\documentclass[10pt,draft]{dis03}
\usepackage{epsf,amsmath}

\newcommand{\as}{$\alpha_s$}
\newcommand{\oas}{$\cal O$($\alpha_s^2$)}

\newcommand{\gev}{\mbox{\,Ge\kern-0.2exV}}
\newcommand{\mev}{\mbox{\,Me\kern-0.2exV}}

\newcommand{\beq}{\begin{equation}}
\newcommand{\eeq}{\end{equation}}

\begin{document}
\title{The present situation in the determination of \as\ 
%Determination of $\alpha_s$ with the DELPHI detector at LEP
}
\author{Oliver Passon \\
Fachbereich Physik, University of Wuppertal\\
Postfach 100 127, 42097 Wuppertal,  Germany\\
E-mail: Oliver.Passon@cern.ch }

\maketitle

\begin{abstract}
\noindent This note reviews the latest measurements of $\alpha_s$ from
event shapes in $e^+e^-$ annihilation at LEP1 and LEP2. 
A critical review of different methods to extract \as\ is offered.
%First the LEP
%QCD working
%group combination of the strong coupling is presented, which is based
%on the so called matched ${\cal{O}}(\alpha_s^2)$+NLLA
%calculations. Additionally we study alternative approaches like
%fixed order calculations with optimized scales and power corrections
%to mean values, which use also  ${\cal{O}}(\alpha_s^2)$ calculations only.  
%Finally we discus the connection between power corrections and the
%renormalisation group invariant (RGI) approach.
\end{abstract}
%%%%%%%%%%%%%%%%%%%%%%%%%%%%%%%%%%%%%%%%%%%%%%%%%%%%%%%%%%%%%%%%%%
\section{Introduction} 
Measurements of the strong coupling \as\ serve as an important 
consistency test of QCD.  The results presented here are based on event shape 
observables in $e^+e^-$ annihilation. 
%These 
%dimensionless quantities characterize the topology of the events, e.g. whether
% the radiation of hard
%gluons gave rise to further jets. 
Since QCD is the theory of  (asymptotically) free quarks 
and gluons, hadronisation effects need to be accounted for. This is
traditionally done  either with phenomenological models (``monte carlo''), 
or with the help of  QCD inspired power corrections. But we will also discuss
results which indicate, that for inclusive mean values non--perturbative
effects are much smaller than originally assumed.

%The QCD calculation  can be performed in different ways as well. The earliest 
%results were based on fixed order perturbation  theory. To extend the 
%applicability into the two jet region  the so called next--to--leading--log 
%approximation (NLLA) was developed. Finally one can combine fixed order
%results with NLLA calculations,  which leads to the matched \oas+NLLA
%prediction. All perturbative QCD calculations suffer from the
%renormalisation scale dependence.

\section{The LEP QCD working group combination}
\begin{figure}[t]
\begin{center}
\begin{minipage}[h]{5.7cm}
\centerline{\epsfxsize=2.0in\epsfbox{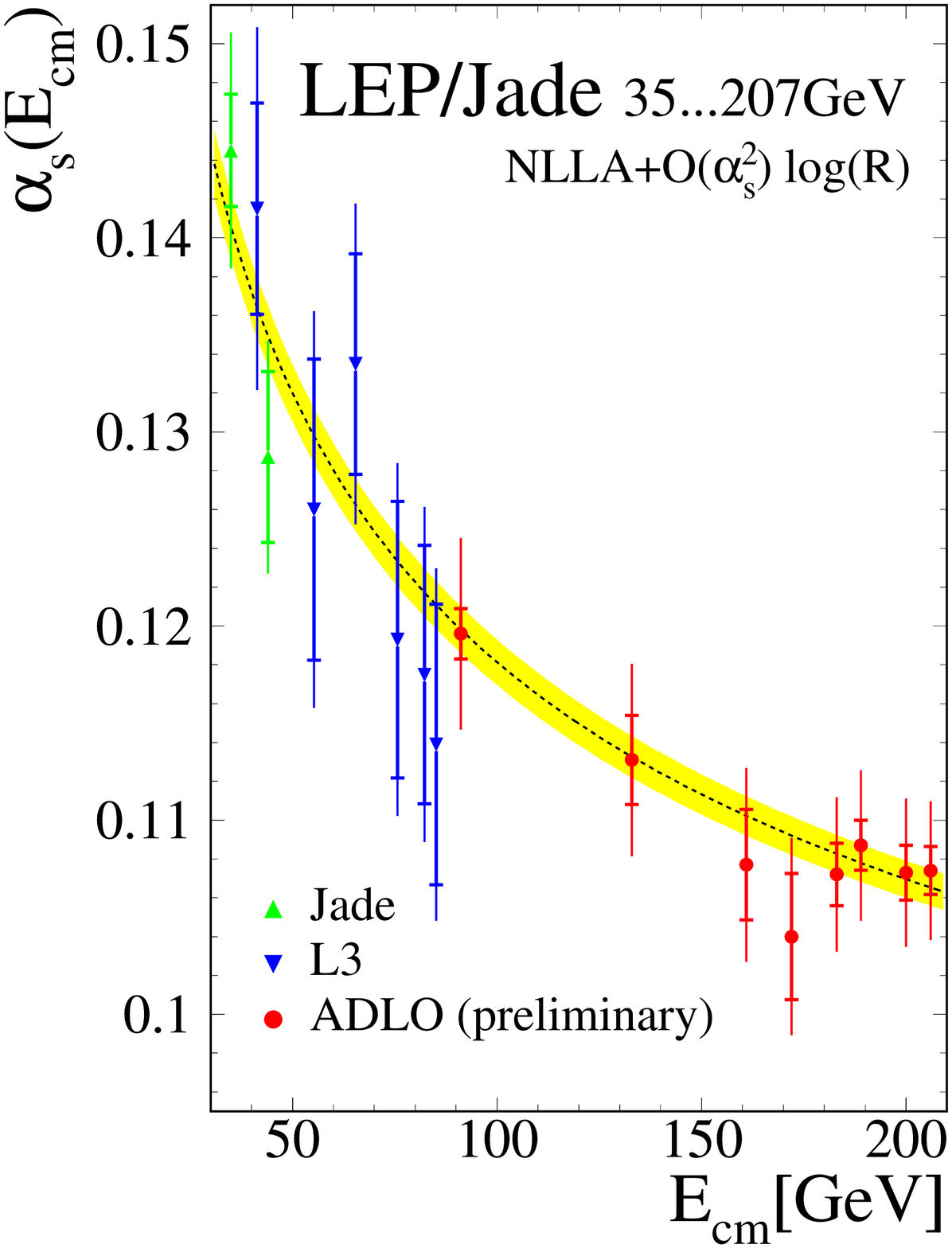}}
\end{minipage}
\begin{minipage}[h]{5.7cm}
\centerline{\epsfxsize=2.5in\epsfbox{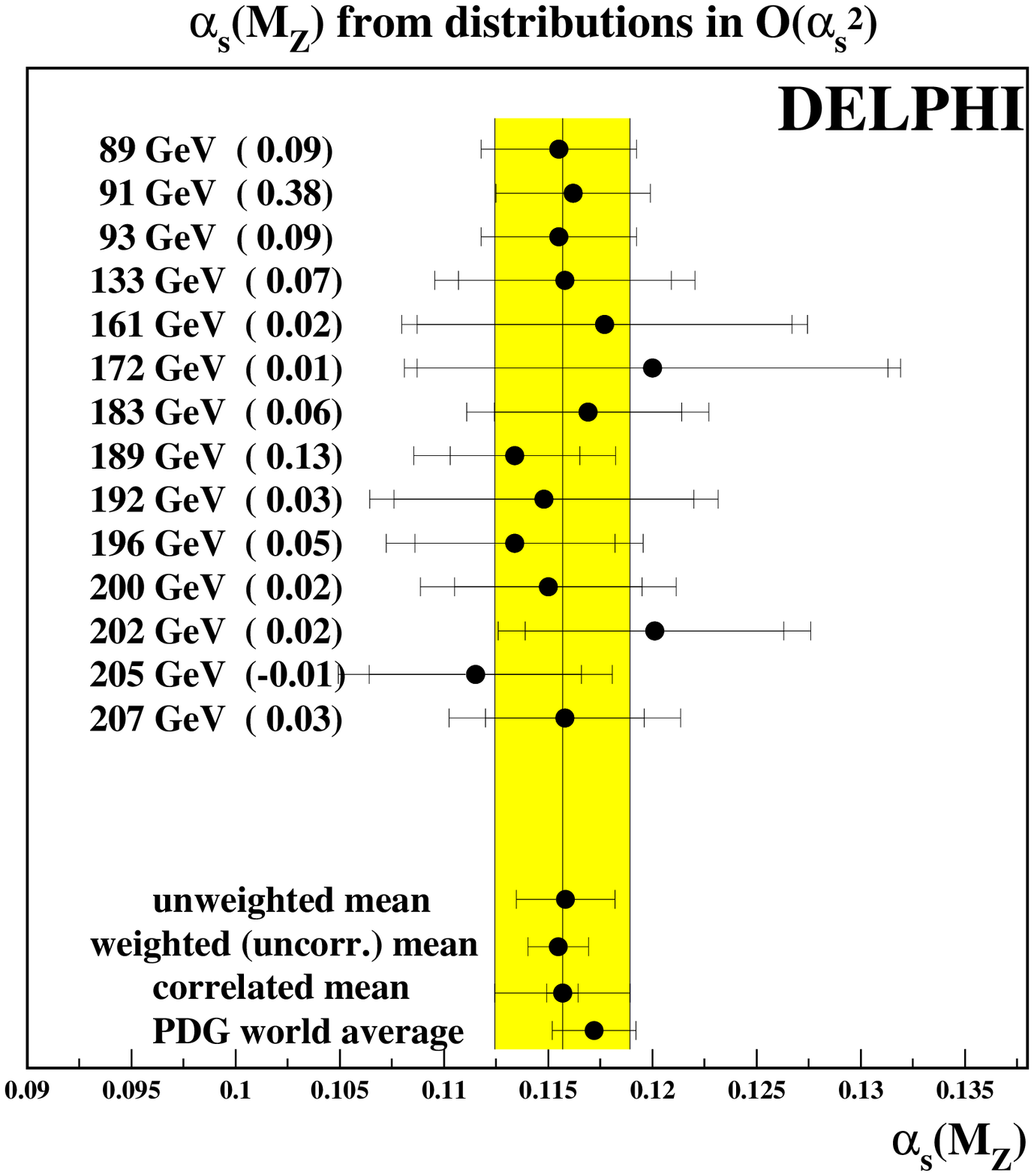}}
\end{minipage}
\caption[*]{\label{wg} {\em Left: Energy dependence of the LEP combined 
$\alpha_s$ values. The band displays the QCD evolution of the mean
  value. Right: DELPHI results on \as\ from distributions at LEP1 and LEP2
  with experimentally optimised scales \cite{mynote}.The inner error bar 
 shows the statistical uncertainty, the outer one the total uncertainty.}}
\end{center}
\vspace{-0.8cm}
\end{figure}

The LEP QCD working group has made serious effort to set up a method
for combining the $\alpha_s$ measurements of the four LEP
collaborations. In the course of this work a common definition and 
implementation of the theoretical predictions was
reached. In contrast to the electroweak working group, not the
 data are combined, but the combination is performed  on the level of
$\alpha_s$ measurements. As input the LEP collaborations
provide the working group with the  $\alpha_s$ fit results based on the
logR matched predictions for several event shape distributions with 
monte carlo hadronisation corrections.
%together with the statistical uncertainties and the results
%when using different hadronisation models. Based on these numbers the
%hadronisation uncertainty can be calculated consistently by the
%working group. Even the theoretical uncertainty is calculated by the
%working group, since it is derived from the theoretical predictions alone. 
%Finally the combination is made, which needs additional assumptions
%regarding the correlation between different experiments, different
%observables and different energies. 
The details of this procedure can
be found in \cite{workinggroup}. The averages over LEP1 and LEP2 data
respectively lead to:
\begin{eqnarray*}
\mbox{LEP1 data:} \quad \alpha_s(m_Z)&=&0.1197 \pm 0.0049 (\mbox{tot}) \\
\mbox{LEP2 data:} \quad \alpha_s(m_Z)&=&0.1196 \pm 0.0046 (\mbox{tot}) 
\end{eqnarray*}
The LEP2 data alone are more precise, since theoretical and
hadronisation uncertainties reduce with increasing energy.
The energy dependence of the \as\ values at LEP1 and LEP2 is shown in 
Fig.\ref{wg} (Left). 
%%%%%%%%%%%%%%%%%%%%%%%%%%%%%%%%%%%%%%%%%%%%%%%%%%%%%%%%%%%%%%%%%%%%%%5
%%%%%%%%%%%%%%%%%%%%%%%%%%%%%%%%%%%%%%%%%%%%%%%%%%%%%%%%%%%%%%%%%%%%%%%
The total error is {\em completely} dominated by the theoretical
uncertainty (i. e. more than 95\%). Traditionally this uncertainty 
was estimated by a variation of the renormalisation scale only. The
exchange with theoreticians within the LEP QCD working group revealed 
additional sources of ambiguity, related to the so--called phase space
condition. Since the NLLA calculations do not 
die out at the phase space boundary $y_{max}$, the replacement
$L=\ln\frac{1}{y} \rightarrow L^{\prime}=\frac{1}{p}\ln \left[ 
\frac{1}{(x_L\cdot y)^{p}}-\frac{1}{(x_L \cdot 
y_{max})^{p}} +1 \right ]$ 
%$L=\ln\frac{1}{y} \rightarrow L^{\prime}=\ln \left[\frac{1}{y}-
%\frac{1}{y_{max}} +1 \right ]$ 
was suggested, with $y$ the observable under investigation. Originally the
choice $x_L=1$ and $p=1$ was made. But in fact other values of $x_L$ and $p$
are formally equivalent i. e. introduce only sub--dominant contributions. This
introduces two more  arbitrary parameters. The working
group now considers the effect on \as\ when changing the
renormalisation scale $\mu$ ($0.5\sqrt{s}\le \mu \le 2\sqrt{s}$), $x_L$ 
($2/3 \le x_L \le 3/2$) and $p$ ($p=1,2$). In order
to avoid double counting the theoretical uncertainty is derived from an
error--band method. For details of the method see
\cite{workinggroup}. Especially the $x_L$ variation leads to an increase in the
theoretical uncertainty. 
%%%%%%%%%%%%%%%%%%%%%%%%%%%%%%%%%%%%%%%%%%%%%%%%%%%%%%%%%%%%%%%%%%%%%%%%%
\section{Alternative approaches}
It is generally assumed that the matched ${\cal{O}}(\alpha_s^2)$+NLLA 
predictions represent the most complete knowledge of perturbative
QCD. But one should bear in mind that the $\chi^2$ of matched fits is
known to be poor which leads to  strong fit--range dependence of \as.
On top of this problems, which were known for long and seemed to be ignored by
part of the community, the appearance of the above mentioned new 
ambiguities makes a look into alternative approaches even more pressing.  
%%%%%%%%%%%%%%%%%%%%%%%%%%%%%%%%%%%%%%%%%%%%%%%%%%%%%%%%%%%%%%%%%%%%%%
\subsection{Experimentally optimised scales}
One of this alternatives is provided by using fixed order calculations
only. It was shown in \cite{siggi}, that this allows a consistent
description of event shape distributions as measured with high precision at
LEP1, provided that the renormalisation scale $\mu$ is treated as an
additional  free parameter in the fit. Hence the resulting
renormalisation  scales are called ``experimentally optimized''. It turns out
\cite{siggi},  that this experimentally optimised scales (EOS) correlate
highly with theoretically suggested scales like the ECH \cite{grunberg} or 
PMS \cite{pms}. This result was confirmed by a study of the
4--jet rate, where the NLO calculation became available only
recently \cite{uwe}. Using this approach for 3--jet like event shape
distributions with the 
DELPHI LEP2 data leads  also to consistent results (see Fig.\ref{wg} (Right)
\cite{mynote}). In all these cases a monte carlo hadronisation correction 
is applied.            

It is a subject of controversial debate weather the EOS procedure is
theoretically founded. As a matter--of--fact the {\em scale} choice for NLO
calculations is in one--to--one correspondence to the choice of a
renormalisation {\em scheme}. In the light of this mathematical property the
scale optimisation can be viewed as the choice of a scheme which
describes the data more properly than the conventional $\overline{MS}$.
The corresponding \as\ values are then retranslated into
$\overline{MS}$  to allow for a direct comparison of the results.

Theoreticians in generally do not like if one confuses ``scales'' and 
``schemes''. It is  claimed that the $\overline{MS}$
scheme choice is merely conventional and that the occurrence of large logs 
(i. e. scales different from $\sqrt{s}$) introduces the need for resummation,
i. e. the NLL approximation. This  argumentation would be more
convincing, if the scale dependence of NLLA calculations would actually
decrease  significantly, which has  {\em not} been observed. 
Additional, the author of this lines is astonished, that a merely 
{\em conventional} choice (like the one for the $\overline{MS}$ scheme) 
is defended with so much vigor, as if it has a deeper meaning nevertheless.    
It should be noted, that another approach to cure the scale dependence
of NLL resummation is offered by the inclusion of renormalon effects 
\cite{einan0}. However, this approach yields smaller values for \as.

It may be suspected, that the refusal of the EOS method is related to its 
attempt to enlarge the regime of {\em perturbative} QCD (this argument applies
even stronger to the RGI method, which will be reviewed later). 
Currently most theoreticians are more attracted by non--perturbative
phenomena. The determination of \as\ is not at the heart of current research
in QCD.  
%%%%%%%%%%%%%%%%%%%%%%%%%%%%%%%%%%%%%%%%%%%%%%%%%%%%%%%%%%%%%%%
\begin{figure}[t]
\begin{center}
\begin{minipage}[h]{5.9cm}
\centerline{\epsfxsize=2.0in
\epsfbox{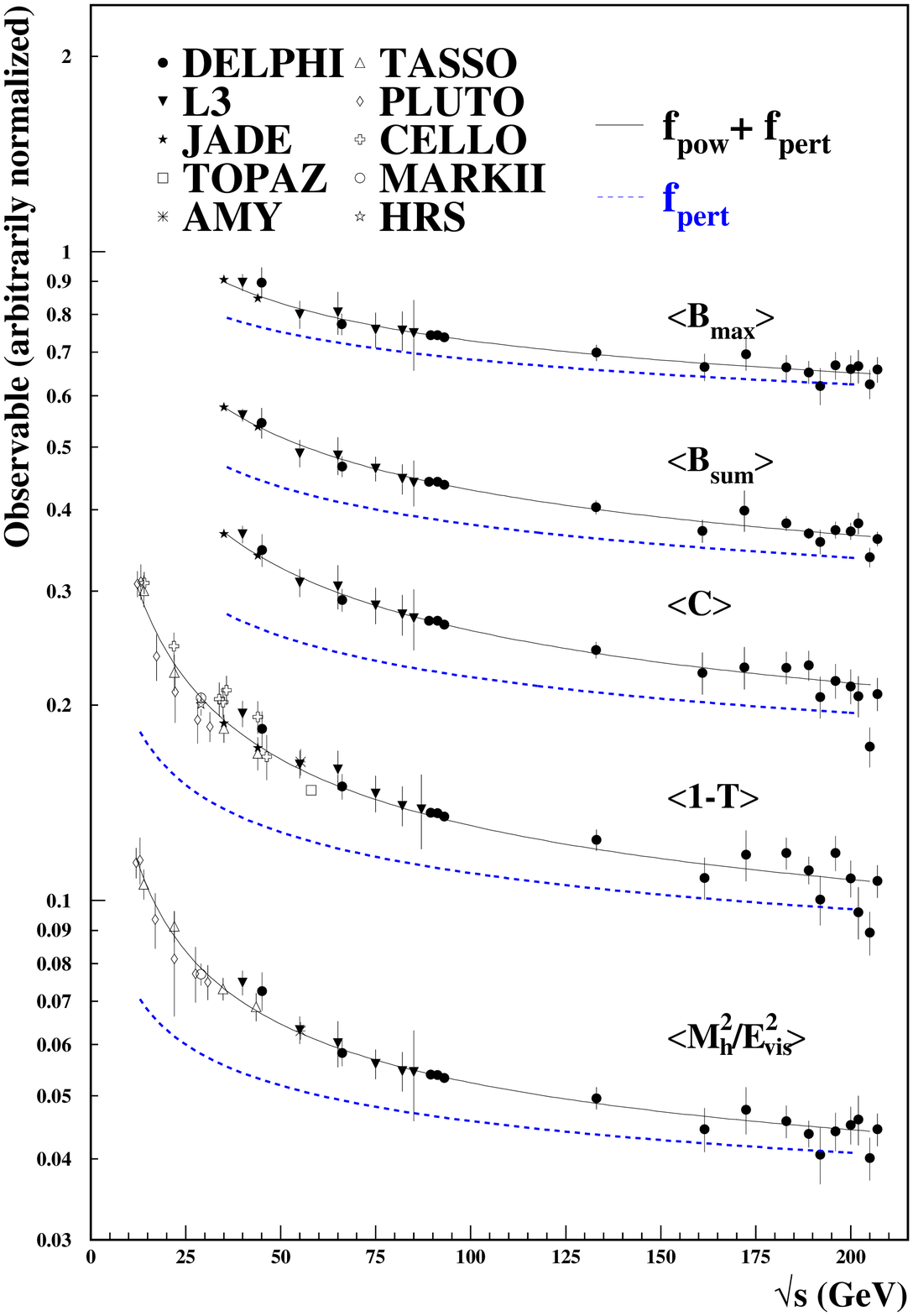}}
\end{minipage}
\begin{minipage}[h]{5.9cm}
\centerline{\epsfxsize=2.0in
\epsfbox{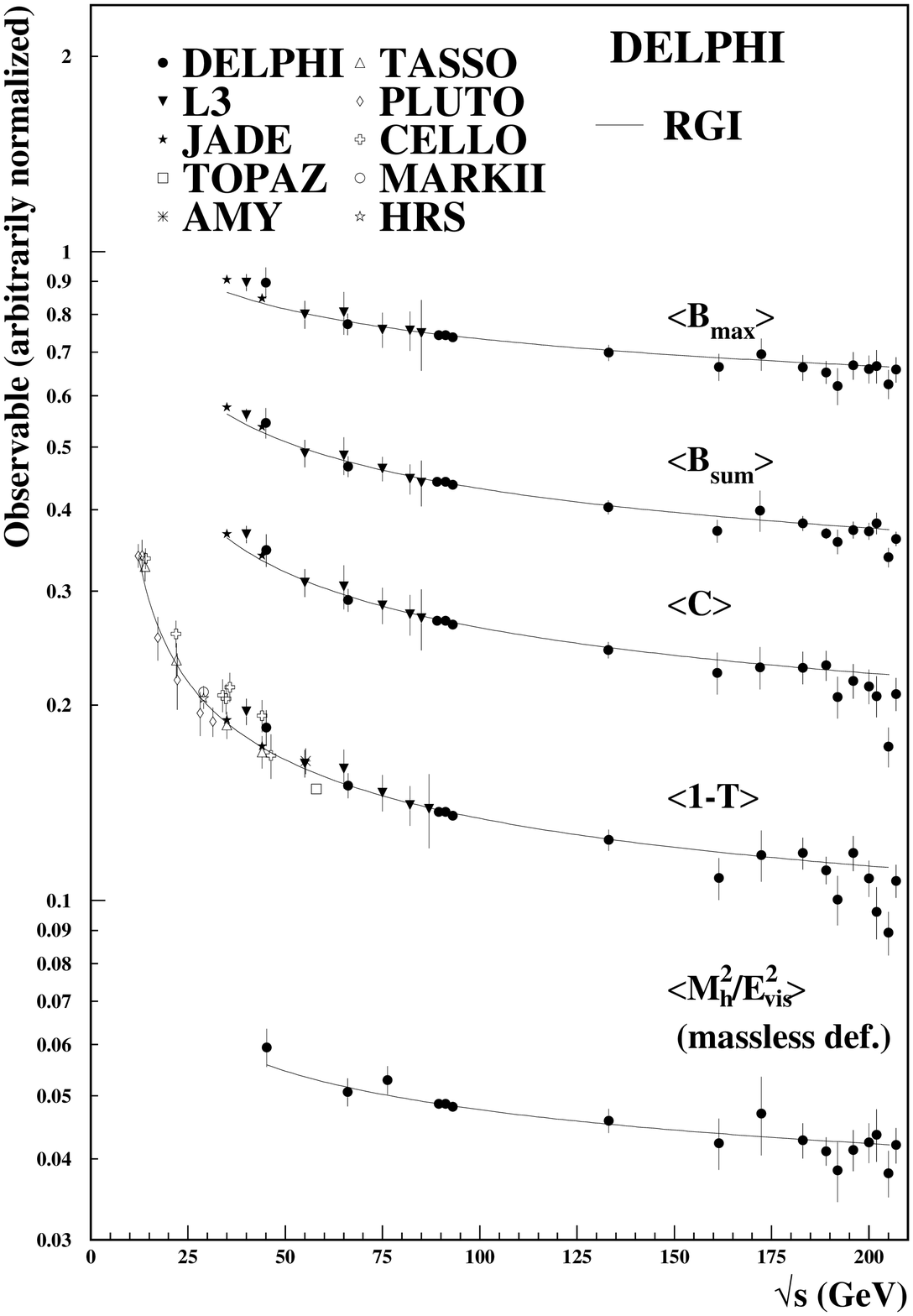}}
\end{minipage}
\caption[*]{\label{r4} 
{\em Left: Fit of the \oas\ + power correction prediction to the energy
  dependence of event shape means. 
Right:  Fit of the purely perturbative RGI prediction
 to the same data set as in the left plot. Due to mass effects
 the heavy jet mass has to be modified.}}
\end{center}
\vspace{-0.8cm}
\end{figure}
%%%%%%%%%%%%%%%%%%%%%%%%%%%%%%%%%%%%%%%%%%%%%%%%%%%%%%%%%%%%%%%%%%%
%%%%%%%%%%%%%%%%%%%%%%%%%%%%%%%%%%%%%%%%%%%%%%%%%%%%%%%%%%%%%%%%%
\subsection{Power corrections}
This leads me directly to the subject of power corrections, one of the prime
examples for the above mentioned new research lines. In this note the ansatz is
applied to mean values of event shapes  \cite{PhysLettB352_451}.The well known
result is displayed in Figure \ref{r4} (left). One gets a proper
description of the  energy dependence of mean values with consistent \as\ and 
$\alpha_0$ values  \cite{mynote} (It should be noted however, that power
corrections to distributions yield systematically lower values for \as\
\cite{ralle}.). Seemingly power corrections provides just a different way to
account for   
hadronisation effects, because they substitute monte carlo
hadronisation corrections. But since power corrections are combined with a
${\cal{O}}(\alpha_s^2)$ calculation, this analysis tells us
also something about scales and the  fuzzy border line between perturbative 
and non--perturbative physics. Since the power
corrections are given to us in the $\overline{MS}$ scheme, the
perturbative part has also to be evaluated in  $\overline{MS}$
(e.g. $\mu=\sqrt{s}$). But the coherent description of the data with
${\cal{O}}(\alpha_s^2)$ calculations in the $\overline{MS}$ scheme seems to
be in striking contradiction to  our earlier claim about the need of
experimentally optimised scales. 
The next section shows how this riddle can be solved.
%Only if the power correction itself would
%be strongly scheme dependent, this riddle could be solved. The following
%section shows, that this happens to be the case.

\subsection{Renormalisation group invariant (RGI) perturbation theory}
Fig.\ref{r4} (Right) \cite{ralle} shows the very same data that was 
displayed in Fig.\ref{r4} (Left). This time the data are compared to the
so--called renormalisation group invariant (RGI) prediction \cite{DG}, which 
is numerically equivalent to the
effective charge scheme \cite{grunberg}. In fact an experimental 
optimisation of the scales would lead to the same curves. The hadronic data are
described by the perturbative calculations only -- there is neither need nor 
room for power corrections. The resulting \as\ values are consistent and close 
to the world average \cite{ralle}. It is of course no surprise that power 
corrections show some scale dependence, but it is perplexing that an 
appropriate  scale can make them to vanish completely. This finding is in 
agreement with the results derived in the context of the renormalon resummation
(see Fig. 5 and 7 in \cite{einan}). 

The other -- and even more important -- virtues of the RGI approach
(especially for a measurement of the  $\beta$ function) can be found in 
\cite{ralle}. 

\section{Summary}
The ability to describe hadronic final states with a perturbative calculation
only (i. e. with one free parameter only) is surprising and its implications 
should be formulated with care. Certainly we do not claim, that 
non--perturbative effects (e.g. hadronisation) do not take place or play no 
role. But for the mean values of event shape distributions  one of the 
two following  statements seems to hold: either their inclusiveness makes them 
essentially insensitive to non--perturbative physics or for some strange
reason  
their effects can be parameterised completely by perturbative QCD, provided a 
proper scheme choice.  In any event it is not justified to choose the \oas\ 
calculation in $\overline{MS}$ while assigning the ``rest'' to  
non--perturbative physics. 
%%%%%%%%%%%%%%%%%%%%%%%%%%%%%%%%%%%%%%%%%%%%%%%%%%%%%%%%%%%%%%%%%%%%%%
\section*{Acknowledgment}
I like to thank Georges Grunberg and especially Einan Gardi for helpful
comments when preparing the final version of this note. 
%%%%%%%%%%%%%%%%%%%%%%%%%%%%%%%%%%%%%%%%%%%%%%%%%%%%%%%%%%%%%%%%%%%%%% 

%%%%%%%%%%%%%%%%%%%%%%%%%%%%%%%%%%%%%%%%%%%%%%%%%%%%%%%%%%%%%%

\begin{thebibliography}{0}
\bibitem{workinggroup} note of the LEP QCD working group in preparation
\vspace{-0.3cm}
\bibitem{siggi} {DELPHI Coll., P. Abreu et al.} {\em E.Phys.J.} {\bf
    C14}(2000)  557. 
\vspace{-0.3cm}
\bibitem{grunberg} G.~Grunberg, {\em Phys. Lett.} {\bf B95} (1980) 70.\\
                   G.~Grunberg,{\em Phys. Rev.} {\bf D29} (1984)  2315.
\vspace{-0.3cm}
\bibitem{pms} P. M. Stevenson, {\em Phys. Rev.} {\bf D23} (1981)  2916. 
\vspace{-0.3cm}
\bibitem{uwe} DELPHI Coll., P. Abreu et al., DELPHI 2001-65 CONF 493 
\vspace{-0.3cm}
\bibitem{mynote} DELPHI Coll., J. Abdallah et al., CERN-EP 327 
\vspace{-0.3cm}
\bibitem{einan0} E. Gardi and J. Rathsman, {\em Nucl.Phys.} {\bf B609} (2001) 
123.\\
 E. Gardi and J. Rathsman, {\em Nucl.Phys.} {\bf B638} (2002) 243.
\vspace{-0.3cm}
\bibitem{PhysLettB352_451} Y.~L. Dokshitzer and B.~R. Webber,
{\em Phys. Lett.} {\bf B352}(1995)  451.
\vspace{-0.3cm}
\bibitem{ralle} {DELPHI Coll., J. Abdallah et al.}, hep-ex/0307048 
\vspace{-0.3cm}
\bibitem{DG} A.~Dhar and V.~Gupta, {\em Phys. Rev.} {\bf D29} (1984)  2822.
\vspace{-0.3cm}
\bibitem{einan} E. Gardi and G. Grunberg, JHEP 9911 (1999) 016.
\end{thebibliography}
\end{document}